\documentclass[showpacs,preprint,jcp,floatfix,superscriptaddress]{revtex4}
\usepackage{ulem}
\usepackage{graphicx}
\usepackage{amssymb}
\usepackage{amsmath}
\usepackage{amsfonts}
\usepackage{caption}
\begin{document}
\title{Out of Equilibrium Characteristics of a Forced Translocating Chain through a Nanopore}
\author{Aniket Bhattacharya}
\altaffiliation[]{
Author to whom the correspondence should be addressed}
\email{aniket@physics.ucf.edu}
\affiliation{Department of Physics, University of Central Florida, Orlando, Florida
32816-2385, USA}
\author{Kurt Binder}
\affiliation{Institut f\"ur Physik, Johannes Gutenberg-Universit\"at Mainz, 
Staudinger Weg 7,55099, Mainz, Germany}
\date{\today}
\begin{abstract}
Polymer translocation through a nano-pore in a thin membrane is
studied using a coarse-grained bead-spring  model and Langevin
dynamics simulation with a particular emphasis to explore out of
equilibrium characteristics of the translocating chain. We analyze
the out of equilibrium chain conformations both at the $cis$ and
the $trans$ side separately either as a function of the
time during the translocation process or as as function of the monomer index $m$
inside the pore. A detailed picture of translocation emerges by
monitoring the center of mass of the translocating chain,
longitudinal  and transverse components of the gyration radii and
the end to end vector.  We observe that polymer configurations at
the $cis$ side are distinctly different from those at the $trans$
side. During the translocation, and immediately afterwards, the
chain is clearly out of equilibrium, as different parts of the
chain are characterized by a series of effective Flory exponents.
We further notice that immediately after the translocation the
last set of beads that have just translocated take a relatively
compact structure compared to the first set of beads that
translocated earlier, and the chain immediately after
translocation is described by an effective Flory exponent $0.45
\pm 0.01$. The analysis of these results is further strengthened
by looking at the conformations of chain segments of equal length
as they cross from the $cis$ to the $trans$ side, We discuss
implications of these results to the theoretical estimates and
numerical simulation studies of the translocation exponent
reported by various groups.
\end{abstract}
\maketitle
\section{Introduction}
Despite a large number of theoretical and numerical studies
voltage driven polymer translocation through a
nanopore\cite{Kasianowitch96,Li01} and its prospective
applications has remained an active field of research with many
open questions. Initial theoretical work of Chuang Kantor and
Kardar\cite{Chuang01}, Kantor and Kardar\cite{Kantor04}, Sung and
Park\cite{Sung96}, Muthukumar\cite{Muthu99}, followed by more
recent theoretical studies by Dubbledam \textit{et
al.}\cite{Dubbledam1,Dubbledam2}, Panja \textit{ et
al.}\cite{Wolterink06,Panja072,Vocks}, Sakaue\cite{Sakaue}, and
Slater \textit{et al.} \cite{Slater_1d} has sparked renewed
interests and numerical work\cite{Milchev04}-\cite{Aniket09}. In a
recent paper\cite{Aniket09} for the case of forced translocation
we have compared predictions from theoretical studies with those
obtained using Langevin dynamics simulation results. We observed
that the driven translocation process is dominated by out of
equilibrium characteristics which may lead to misinterpretation of
the time averaged data for the translocated chain. We further
noticed that one needs to look at the chain segments at the $cis$
and the $trans$ side separately to get a true picture of the
translocation process.
\par 
The purpose of our work is to collect the central facts that show  the
nonequilibrium character of forced translocation in its various
aspects. Of course, development of a new theory
that puts all our observations in a coherent framework
would be very desirable, but is beyond the scope of our work.
However, we present some elements of a phenomenological theoretical description
of nonequilibrium behavior, in terms of an effective time-dependent
correlation length that describes on which scales excluded volume
interactions during translocation are equilibrated.
We believe that a coherent description of these
aspects is provided by our paper, and thus it will constitute
a sound basis on which the future development of such a
desirable theory can be based. \par
To explicitly demonstrate the out of  
equilibrium characteristics of relevant physical parameters are studied 
in detail by
monitoring the $trans$ and the $cis$ characteristics separately as
a function of the chain segments separately at the $cis$ and  $trans$
compartments. 
To emphasize the role of out of equilibrium aspects
we have taken a \textit{reductio ad absurdum} approach. We first
hypothesize that the chain undergoing forced translocation is at
each instant of time in equilibrium; we then demonstrate violation
of this hypothesis for the forced translocation process in several
physical quantities obtained from simulation studies. Of course,
we also have carried out simulation studies for the unbiased
translocation so that these results can serve as reference for the
specific pore-wall geometry that we choose. In the next section we
describe the model. In section III we present the results from
Langevin dynamics simulation. In section IV we discuss in detail
the relevance of these results
\par
\section{THE MODEL}
We have used the ``Kremer-Grest'' bead spring model to mimic a strand of
DNA \cite{Grest1}.
Excluded volume interaction between monomers is
modeled by a short range repulsive LJ potential
\begin{eqnarray*}
U_{LJ}(r)&=&4\epsilon [{(\frac{\sigma}{r})}^{12}-{(\frac{\sigma}
{r})}^6]+\epsilon \;\mathrm{for~~} r\le 2^{1/6}\sigma \\
        &=& 0 \;\mathrm{for~~} r >  2^{1/6}\sigma\;.
\end{eqnarray*}
Here, $\sigma$ is the effective diameter of a monomer, and
$\epsilon$ is the strength of the potential. The connectivity between
neighboring monomers is modeled as a Finite Extension Nonlinear
Elastic (FENE) spring with
\[ U_{FENE}(r)=-\frac{1}{2}kR_0^2\ln(1-r^2/R_0^2)\;,\]
where $r$ is the distance
between consecutive monomers, $k$ is the spring constant and $R_0$
is the maximum allowed separation between connected monomers.
We use the Langevin dynamics with the equation of motion
\[ \ddot { \vec{r}} _i = - \, \vec
{\nabla} U _ i - \Gamma \dot { \vec{r}} _i \, + \vec {W} _ i (t) \;.\]
Here $\Gamma$ is the monomer friction coefficient and
$\vec{W} _ i (t)$, is a Gaussian white noise with zero mean at a temperature T, and
satisfies the fluctuation-dissipation relation:
\[ < \, \vec{W} _ i (t)
\cdot \vec{W} _ j (t') \, > = 6k_BT \Gamma \, \delta _{ij} \, \delta (t
- t ')\;.\]
The purely repulsive wall consists of one monolayer of immobile LJ particles of diameter 1.5$\sigma$
on a \textit{triangular lattice}
at the $xy$ plane at $z=0$. The pore is created by removing the particle at the center.
Inside the pore, the polymer beads
experience a constant force $\mathbf{F}=F\hat{z}$  along positive $z$ direction ($F\hat{z} \ne 0$ for $|z| \le \sigma/2$) and a repulsive potential from the inside wall of the pore.
The reduced units of length, time and temperature are chosen to be  $\sigma$,
$\sigma\sqrt{\frac{m}{\epsilon}}$, and $\epsilon/k_B$ respectively.
For the spring potential we have chosen $k=30$ and $R_{ij}=1.5\sigma$, the friction coefficient
$\Gamma = 1.0$, and the temperature is kept at $1.5/k_B$ throughout the simulation. Evidently 
with an applied bias translocation occurs along $+z$ direction while for the unbiased case the 
translocation can occur along $\pm z$ directions.  \par

For the case of unbiased translocation we have chosen chain lengths $N$ = 17, 33, 65 ,129, and 257
and put the middle monomer symmetrically inside the pore ($z=0$) with equal number of monomers
at the $cis$ and the $trans$ side and let the translocation occur in either $+z$ or $-z$ directions.
We have checked that for a large number of repeated trials the probability for translocation from
right to left or vice versa is the same. In this case each chain is equilibrated keeping the
middle monomer at the fixed location $z=0$ inside the pore and then releasing it for translocation.
The results are averaged over at least for 2000 trials.

For the case of forced translocation we carried out simulations
for chain lengths $N$ from $8 - 256$ for several choices of the
biasing force $F = 2$, $4$, $6$, and $12$ respectively. For most
cases we show the results for $F=6$. Initially the first monomer
of the chain is placed at the entry of the pore. Keeping the first
monomer in its original position the rest of the chain located at 
the $cis$ compartment is then
equilibrated for times at least an amount proportional to the
$N^{1+2\nu}$. The chain is then allowed to move through the pore along 
$z$ direction. Even for the forced translocation for short times back and forth motion 
of the chain is possible. However, a
successful translocation occurs when the entire chain initially located at the 
$cis$ side moves to the $trans$ side   
driven by the force $\mathbf{F}=F\hat{z}$ present inside the pore. When the last monomer
exits the pore we stop the simulation and note the translocation
time and then repeat the same for $5000$ such trials.  In
comparison to reality, our model is drastically simplified: we
deal with a neutral homopolymer without explicit solvent, there
are neither hydrodynamic interactions nor long range electrostatic
forces. However, our model is of the same type as most of the
currently available theoretical work. We feel that this simple
limiting case needs to be understood first, before more realistic
cases may be tackled. \par

\section{Results}
As mentioned in the introduction that along with the results for
the forced translocation we also present results for the unbiased
case, as necessary for comparison. Except for Fig.~\ref{cmznh_N}
where plots are shown for four different biases the remaining
figures for the forced translocation refer to $F=6.0$. During the translocation process 
we have monitored several quantities. Before we present our results here we explain the 
notations used. The transverse and longitudinal components of the gyration radii are 
denoted as $R_{gt}$ and $R_{gl}$ respectively. To study various aspects of the chain 
segments residing at the $trans$ and $cis$ sides we characterize the chain monomers with 
the index $m$, such that $m=1$ and $M=N$ denote the first and last beads respectively. 
In the case of forced translocation initially the first bead($m=1$) is placed inside the pore 
and translocation ends when the last bead ($m=N$) leaves the pore. In addition, we denote the 
number of monomers $n$ residing either at the $cis$ or the $trans$ side of the pore 
and study several quantities as a function of the number of monomers $n$. For example,  
$R_{gt}^{cis}(n)$ and $R_{gl}^{cis}(n)$ 
denote the transverse and longitudinal components of the gyration radii for 
a segment of length $n$ residing at the $cis$ side. 
Likewise, $R_{gt}^{trans}(n)$ and $R_{gl}^{trans}(n)$ 
represents the transverse and longitudinal components of the gyration radii 
for a segment of length $n$ residing at the $trans$ side respectively. 
During the simulation along with these gyration radii we 
have monitored the mean first passage time (MFPT),
the end-to-end vector $R_N$, location of of the $z$ coordinate of the 
middle monomer $z(N/2)$, and the location of the 
$z$-coordinate of the center of masses of
the segments residing at the $cis$ ($Z_{CM}^{cis}(n)$) and the $trans$ 
($Z_{CM}^{trans}(n)$) sides respectively. In each
case for a quantity $A$ the notation $\langle A \rangle$ refers to
the average over 2000-5000 independent runs.
\subsection{Translocation exponent for the unbiased translocation}
First, as a reference we have checked the scaling exponents for
the unbiased translocation for the choice of parameters used here.
This is shown in Fig.~\ref{tauf0} where we find $\langle \tau
\rangle \sim N^{2.2}$ and $\langle R_g \rangle \sim N^{0.61}$.
Therefore we verify that for unbiased translocation the
translocation exponent $\alpha$ ($\langle \tau \rangle \sim
N^\alpha$) scales as $\langle \tau \rangle \sim N^{1+2\nu}$ as
previously found by us\cite{Milchev04} in a slightly different
context using Monte Carlo (MC) and others in
2D\cite{Kaifu06a,Wei07}.
\subsection{Mean first passage time (MFPT)}
Fig.~\ref{mfpt} shows the MFPT time for a forced translocating
chain as a function of the translocated segment $m$ for several
chain lengths. We include this curve for future discussions as one
readily  gets the average translocation time $\langle \tau (m)
\rangle $ of a particular monomer index $m$ or vice versa. 
Please note that for forced translocation at the beginning and end of the 
translocation the monomer index has a value $m=1$ and $m=N$ respectively, therefore, a large value 
of $m$ relates to a relatively a late stage of the translation process and vice versa. 
For
larger chain lengths the non-linear nature in the form of an
\textit{S} shaped curve becomes more prominent where we notice
that the curvature is negative (concave) for $ m < N/2$ and
becomes positive for $ m > N/2$. The last few beads take
relatively shorter time to translocate.
\subsection{What is the location of $Z_{CM}(N/2)$ ?}
If the translocating chain were in local equilibrium then one
would expect that when the chain is half way through the
translocation process the position of the $z$ component of the
center of mass $Z_{CM}(N/2)$ should be located symmetrically with
respect to the $cis$ and the $trans$ side. This should correspond
to $z=0$ for our choice of coordinate system. Fig.~\ref{cmznh_t}
shows the distribution of the location of $Z_{CM}(N/2)$ and the
average value $\langle Z_{CM}(N/2) \rangle $ for three different
chain lengths. Please note that while calculating  $\langle Z_{CM}(N/2) \rangle $
we have taken into consideration the back and forth motion of the chain. Therefore, 
here  $\langle Z_{CM}(N/2) \rangle $ implies, first, to determine the average 
of $Z_{CM}(N/2)$ from a single run considering all the events when the monomer 
index $m = N/2$ inside the pore and then taking average over all different runs. 
We note that with increasing chain length the
location of  $Z_{CM}(N/2)$ goes deeper  in the $cis$ side of the
chain. When we plot $\langle Z_{CM}(N/2) \rangle $ as a function
of the chain length $N$ shown in Fig.~\ref{cmznh_N} it appears that 
after a certain \textit{crossover length} the distance increases
approximately linearly with the chain  length for larger chains.
It also appears that this crossover length
increases as the bias becomes small. This implies that finite
chain lengths effects become progressively smaller as a function
of the increasing bias, a result that seems counter-intuitive. However, with 
the limited data available at this time we do not make a strong claim, but indicate 
to such a trend. In order to establish more accurately the exact location of a possible 
crossover, it will require runs with many more values of bias and longer chains making  
the simulations prohibitively long.  
\subsection{Behavior of $Z_{CM}^{cis}(n)$ and  $Z_{CM}^{trans}(n)$}
This asymmetry in the location of  $\langle Z_{CM}(N/2) \rangle $
becomes more obvious when we look at the absolute values of 
positions of the center
of masses on the $cis$ side $|Z_{CM}^{cis}(n)|$  and the $trans$
side $|Z_{CM}^{trans}(n)|$ separately. Again, if the local
equilibrium condition should hold, one would expect that as a
function of the $trans$ and $cis$ segments the behavior of
$|Z_{CM}^{trans}(n)|$ and $|Z_{CM}^{cis}(n)|$ should be the same. For
comparison, and as a reference we first show the plot for the case
of unbiased translocation in Fig.~\ref{cmzs_unbiased} where we
notice that the longitudinal and the transverse components of the
gyration radii for both the $cis$ and $trans$ segments are almost
indistinguishable, as anticipated. We further note that for large
portion of the chain except for very small and large values of the
monomer index $\langle |Z_{CM}^{cis,trans}|\rangle \sim m^{0.56}$.
If the configurations of the chains in the case of unbiased
translocation just realize a sequence of equilibrium
configurations, we expect $\langle|Z _{CM}^{cis, trans}|\rangle \sim
m^\nu$ with $\nu \approx 0.59$ the equilibrium Flory exponent,
when $m \rightarrow \infty$. Although the effective exponent
$(0.56)$ is somewhat smaller, it is clear that in this case the
translocating chains are close to equilibrium. This should be
contrasted with the corresponding case for the forced
translocation where we observe a very different behavior as shown
in Fig.~\ref{cmzs_forced} for the $cis$ and the $trans$ segments.
Since the $\langle Z_{CM}^{cis}(n) \rangle$ takes negative values at the 
$cis$ side we consider the absolute values $\langle |Z_{CM}^{cis}(n)| \rangle$ 
and $\langle |Z_{CM}^{trans}(n)| \rangle$. 
Using Fig.~\ref{mfpt} one can translate the figure from the
variable $m$ to the corresponding MFPT. One notices that the $cis$
component increases almost linearly for $ 0 < m < N/2$ and scales
as $\langle |Z_{CM}^{cis}(n) |\rangle \sim m^{0.91}$, whereas the
$trans$ part after a linear rise for small $m$, increases at a
much slower rate compared to the $cis$ part. We further notice
$\langle |Z_{CM}^{trans}(n)| \rangle \sim m^{0.45}$, the value of
the slope (0.45) in this case exactly half of the corresponding
value of the $cis$ part. We notice similar qualitative behavior
for $\langle R_{g}^{cis}(n) \rangle$ and $\langle
R_{g}^{transs}(n) \rangle$ and defer the physical explanation of
this behavior to subsection-F.
\subsection{Gyration radii: time dependence}
To get a better idea of the out of equilibrium characteristics of
the forced translocating chain we have compared the time
dependence of gyration radii with those for the unbiased chain
shown in Fig.~\ref{rgt}. For unbiased translocation $\langle
R_{gl} \rangle > \langle R_{gt} \rangle$ is due to the presence of
the wall that breaks the isotropy symmetry and gyration radius
along the translocation direction becomes larger when calculated
for the entire chain. This anisotropy occurs already for a polymer
chain anchoring with a chain end at a flat impenetrable surface (a
``polymer mushroom''). For unbiased translocation, the typical
chain conformations are those of equilibrium mushrooms containing
$m$ and $N-m$ segments, respectively. When we compare these time
dependencies with those for the chain undergoing forced
translocation we notice that the forced chain undergoes
considerable variation in its size. However, despite this large
variation in size when we calculate the average gyration radii we
find $\langle R_g \rangle \sim N^\nu$\cite{Aniket09}. Therefore,
if one does not plot and see this explicit time dependence, one
would tend to make an assumption that only one bead of the chain
is driven inside the pore, and therefore the chain is still
described by the equilibrium Flory exponent\cite{Kantor04}. The
argument that immediately comes to ones mind that the variations
are cancelled (Fig.~\ref{rgt}(b)) which we find is wrong and it is
inappropriate to describe the chain by a single exponent. In the
next subsection we will provide the correct physical reason why
the chain is still described by the equilibrium Flory exponent.

\subsection{Gyration radii components at the $cis$ and $trans$ side}
To further demonstrate the out of equilibrium aspects of the
driven translocating chain we have monitored the time dependence of
the transverse and longitudinal components of the $cis$ and
$trans$ part of the gyration radii separately and compared these
characteristics with those for the unbiased translocation.
Fig.~\ref{rgs_unbiased} shows the plots of the transverse and
longitudinal components for the $cis$ side 
($\langle R_{gt}^{cis}(n) \rangle$, $\langle R_{gl}^{cis}(n) \rangle$) and
for the $trans$ side ($\langle R_{gt}^{trans}(n) \rangle$,
$\langle R_{gl}^{trans}(n)$) separately for the unbiased
translocation. As expected, except for the few first and last
beads (which may depend on the local condition) the conformations
for the $cis$ and $trans$ parts are the same\cite{comment1}. We
further notice that for small values of segments $m$ $R_{gl} >
R_{gt}$; however this difference might arise due to finite chain
length that vanishes for large values of $m$. Within our error
limits, these data are compatible with a scaling of all these
radii according to a law $R\propto N^\nu$ with $\nu \propto 0.59$
being the expected Flory exponent.\par 
For the case of forced
translocation (Fig.~9) one immediately notices that the $\langle
R_g(n) \rangle $ of the chain on the $cis$ and $trans$ sides are
described by different effective Flory exponents. The effect is
most pronounced for the longitudinal component. We have calculated
the slopes for $N/8 < m < N/4 $ and find that $ \langle
R_{gl}^{trans}(n) \rangle \sim m^{0.6}$ and
 $\langle R_{gl}^{cis}(n) \rangle \sim m^{0.8}$. This result is consistent with the recent theory proposed by
Sakaue\cite{Sakaue} based on the propagation of tension along the
chain. During the forced translocation process at early time ($m
<< N$) the translocated monomers relax faster and are described
roughly by the equilibrium Flory exponent. However, during the
short time needed that a small number of $n$ monomers are pulled
through the pore by the force, the remaining $N-n$ segments do not
have enough time to equilibrate their configuration.  One can
think of this effect as the size of $R_g^{cis}(N-n)$ remaining
the \textit{same as} $R_g^{cis}(N)$  but with a few $less$ number of
monomers because of those that have translocated. This leads to 
a very weak dependence of  $R_g^{cis}(n)$ for $n \sim N$ (a saturation 
observed in Fig.~9) and makes the
effective Flory exponent higher on the $cis$ side at the early
stage of the translocation process. As time proceeds the number of
monomers on the $cis$ side also decreases causing the slope 
(Fig. 9) also to decrease). This difference in behavior of
$R_{g}^{trans}(n)$ and $R_{g}^{cis}(n)$ as manifested in our
simulation studies should serve as useful information for the
development of detailed analytical theories of forced polymer
translocation through a nano pore. Since the translocation time in
the case of forced translocation for large $N$ is much smaller
than the Rouse time needed to equilibrate a chain, the
conformations of the chains that just have translocated are much
too compact (Fig.~10), leading to a scaling $R_g^{trans} \sim
N^{\nu_{eff}}$ with $\nu_{eff} \sim 0.5$. \par
\subsection{Analysis of the subchain relaxation on $cis$ and $trans$ sides}
In order to get a better picture of the chain relaxation at the
$trans$ and $cis$ side we have further investigated the relaxation
of sub-chain conformations during the translocation process. Since
this analysis requires a relatively longer chains we will restrict
our discussion for chain of lengths $N=128$ and 256 consisting of 
8 subchains each of length 16 (for $N=128$) or 32(for $N=256$) respectively. 
For the rest of the
discussion in this section let's consider the chain length $N=256$
as the qualitative features for the $N=128$ chain 
with 8 subchains each consisting of 16 monomers are very similar. Let's denote these
subchains of length 32 as $n_i$ ($i = 1,2,\cdot\cdot 8$).
Therefore the $i-th$ chain segment $n_i$ consists of monomers
having indices from $(i-1)*32+1$ to $i*32$. At the start of the
translocation process initially, all subchains are in equilibrium,
and have their values for their end-to end distance and gyration
radius components as appropriate for a calculation done for the
mushroom in equilibrium, which is not yet translocating and the
entire chain being located at the $cis$ side. Once the
translocation starts, we study all these subchains at times when
the last monomer of each segment $n_i$ has just translocated from
the $cis$ to the $trans$ side. Therefore, for the chain length
N=256 consisting of 8 subchains, at the end of a successful
translocation process we will have 8 different subchain
conformations starting with $n_{trans}=1$ and $n_{cis}=7$ until
all the segments are translocated when $n_{trans}=8$ and
$n_{cis}=0$. We then analyze these subchain conformations averaged
over 2000 repeated trials.\par It is expected that the subchain
$n_i$ which has just translocated and its two neighboring
subchains $n_{i-1}$ and $n_{i+1}$ will be affected by the driving
force and out of equilibrium characteristics will dominate most.
But the chain segments which are further away will progressively
be closer to equilibrium. In other words, by the time the subchain
$n_i$ has just translocated, the translocated chain segments
$n_1,n_2,n_3 \cdot \cdot n_{i-2}$ will have more time to relax (as
a function of the decreasing segment index $i$) compared to the
segment $n_{i-1}$, while the segments $n_{8}$, $n_{7} \cdot \cdot$
will remain mostly in equilibrium and hardly feel that the
subchain $n_i$ has just translocated. Thus if we look at the
conformations for all the segments at these times we expect to see
that the $cis$-subchain closest to the pore is mostly stretched
with a higher value of the Flory exponent, while the
$trans$-subchain which just finished translocation is compressed
most with a lower value of the Flory exponent. These higher(lower)
value at the $cis(trans)$ side will progressively taper towards
the equilibrium value on either side of the pore. The increased
value of the Flory exponent will appear like a propagating defect.
\par In the accompanying Fig.~\ref{subchain32} (subchain length 32) 
we have plotted the gyration radii and the
corresponding transverse and the longitudinal components for the
chain segments at times immediately after subchains $1,2,\cdot
\cdot 8$ have translocated. 
We have carried out the same analysis (not shown here) for a chain length 
$N=128$ for subchain lengths 16. This enables us to extract the effective Flory 
exponents $\nu_{eff}$ for each point in Fig.~\ref{subchain32} calculated from the gyration 
radii for subchain lengths 32 (for $N=256$) and 16 (for $N=128$) respectively. This 
is shown in Fig.~\ref{subchain_nu}. 
In Figs.~\ref{subchain32} and 
~\ref{subchain_nu} we kept the vertical scale the same for all the
graphs for ready comparison of the gyration radii and the $\nu_{eff}$ at different
stage of the translocation process. First, one notices that after
the translocation process, the chain is still compressed.
Secondly, at the $trans$ side the slope of the line connecting the
points increases monotonically (not shown) from the left to the
right, the gyration radii of the $trans$-segment closest to the
pore being the least (such as in Fig.~\ref{subchain32}(f) and
Fig.~\ref{subchain32}(g)). This is due the fact, as stated above,
that the farthest $trans$-segment have more time to approach
equilibrium configuration from its initial compressed state at the
entry of trans side. Likewise, on the $cis$ side the slope of the curve
joining the $cis$ segments decreased monotonically for the same
reason that the farthest segment is still in equilibrium and the
$cis$-segment closest to the pore is mostly stretched. 
Finally, from Fig.~\ref{subchain_nu} we indeed notice that the effective 
Flory exponent is large near the pore and the non-monotonic behavior 
looks like a propagating defect. 
It is clear
from the data that subchain relaxation proceeds exactly the way as
contemplated in above.\par
\subsection{Properties of a translocated chain}
In order to interpret the properties of a chain which has just
translocated, we introduce the concept of a time-dependent
correlation length $\xi(t)$ describing the range over which (on
the $trans$ side of the membrane) excluded volume effects are
established (at a time $t$ after the translocation process has
begun). We start from the fact that for the Rouse model with
excluded volume interactions the length (chain radius $R$) and
relaxation time $\tau$ are related as $\tau \propto R^z\propto
N^{\nu z} = N^{2 \nu +1}$ where the dynamic exponent $z=2+1/\nu$.
 Consider now the ``Gedankenexperiment'' that for a Gaussian chain
at time $t=0$ the excluded volume interactions are suddenly
switched on. Then the chain starts to swell, its radius will grow
from the Gaussian size $(R\propto N^{1/2})$ to the swollen size
$(R\propto N^\nu$ with $\nu \approx 0.59$) during the time $\tau$.
At time $t$, one can envisage this gradual swelling that on a
length scale $\xi(t)$ excluded volume has already been
established, but not on scales much larger than $\xi(t)$; so the
chain is a Gaussian string of blobs of size $\xi(t)$. One expects
that $\xi(t) \propto t^{1/z}$ and then indeed there is a smooth
crossover to the equilibrium size $R \propto N^\nu$ at $t=\tau$,
with $\xi (\tau)\propto N^\nu$.\\
In a forced translocation process, the monomers move to the
$trans$ side so fast that the equilibrium structure cannot develop
on the scale of the full chain, but only the first $N_{eq}$
monomers that have passed the pore can develop in their
configuration the proper correlations due to excluded volume. We
can estimate (for large enough $F$) this number $N_{eq}$  as follows.
Following the same argument for the unbiased translocation we now need 
to replace $N$ by $N_{eq}$ so that the correlation length $\xi(\tau)$ 
at time $\tau$ is given by
$\xi(\tau) \sim N_{eq}^\nu$ where that translocation time $\tau \sim N^\alpha/F$.
\begin{eqnarray*} 
\therefore N_{eq} \sim \left[\xi(\tau)\right]^{1/\nu} \sim \tau^{\nu/z} 
\sim \left(\frac{N^\alpha}{F}\right)^{\frac{1}{\nu z}}.
\end{eqnarray*}
Substituting $z = 2 + 1/\nu$ in above equation one gets 
\begin{eqnarray}
\label{eq1}
N_{eq} \sim \xi(\tau)^{1/\nu} \sim N^{\alpha /(2\nu+1)} F^{-1/(2\nu+1)}.
\end{eqnarray}
Therefore, the corresponding size $R$ is 
\begin{equation}\label{eq2}
R \sim N_{eq}^\nu \sim \xi(\tau) \propto N^{\alpha \nu/(2 \nu+1)} F^{-1/(2 \nu +1)}
\end{equation}
For $\alpha = 1.36$ [28] we obtain $\alpha \nu/(2 \nu +1)=0.37$
while the value for $\alpha$ proposed in [4], $\alpha =1+\nu$
would yield $\alpha \nu/(2\nu +1)=0.43$. Of course, the estimate
in Eq.~\ref{eq2}, based on the $N_{eq}$ monomers that had enough
time to equilibrate during the translocation time, is only a lower
bound for the actual radius $R_{trans} $ of the translocated
chain, because the additional $N-N_{eq}$ monomers (translocated,
but not yet equilibrated) can lead to some further increase in an
uncorrelated way (so $R\propto N^{1/2}$ probably is an upper
bound). However, it remains a challenge to clarify whether the
effective exponent $\nu_{eff} \approx 0.45$ observed for several
properties on the $trans$ side (Fig. 6, Fig. 10) really describes
the asymptotic regime or is still affected by some correction to
scaling effects; recall that in the data for unbiased
translocation effective exponents vary from $\nu_{eff} \approx
0.55$ (Fig.~5) to $\nu_{eff} \approx 0.63$ (fig.~1) around the
true asymptotic value $\nu \approx 0.59$.

Another problem that is not understood is to explain the rather
different behavior of $R_{g \ell}$ and $R_{gt}$ for large values
of $m$ (close to $N$) in Fig.~14. These data are presumably
dominated by the $N-N_{eq}$ monomers of the translocated chain
that are still far from equilibrium. Therefore it is clear that
the evolution of the translocated chain configuration with time
(or with $m$, respectively) cannot be described in a simple
manner.

Finally, it would be interesting to clarify the crossover between
forced and unbiased translocation as $F \rightarrow 0$. Note that
Eqs.~\ref{eq1}, \ref{eq2} do not apply in this limit. Rather a
simple scaling assumption could be

\begin{equation}\label{eq3}
\langle \tau \rangle = N^{2 \nu +1} f(FN^{2\nu +1 - \alpha})
\end{equation}
where the scaling function $f(\zeta)$ behaves as $f(0)= const$
(unbiased translocation) while $f(\zeta \gg 1)\propto 1/\zeta$ in
order to recover Eq.~\ref{eq1} in the limit of large enough $F$.
Fig.~4 is a first indication that in the structure of the chain
there is indeed an interesting and nontrivial dependence on $F$,
but clearly more work is needed to fully clarify the aspects of
this behavior by extensive simulations, which are very
time-consuming and therefore beyond the scope of the present work.
But a safe general conclusion is that the theory of 
forced translocation needs to
adequately account for the non-equilibrium aspects of the
translocation process and the resulting strong asymmetry of the
properties of the \textit{cis} and \textit{trans} parts of the
chain.
\subsection{Distribution of the end to end length}
To get a better idea of the out of equilibrium aspects of the translocating chain we have
looked at the distribution of the transverse and longitudinal components of the end to end
vector $\mathbf{R}$ for the chain which has just translocated shown in Fig.~\ref{r1n_hist}. For
comparison we have also shown the corresponding distribution of the
equilibrated chain configurations located at the $cis$ side at time $t=0$
at the beginning of each translocation process. One immediately notices that the
the distribution of $\mathbf{R}$ is much more restricted
compared to its equilibrium configuration.\par

\section{Summary and Discussion}
In summary, several aspects of forced polymer translocation are
studied in this paper which explicitly point towards the out of
equilibrium aspects of the translocation process. In several cases
we also furnish data for the unbiased translocation for identical
parameters of the simulation. From the series of data one could
say that immediately  after the translocation starts till it ends,
the entire chain is always out of equilibrium. However, the nature
and extent of this out of equilibrium aspect is different at the
$cis$ and $trans$ sides and evolves with time. The ``subchain''
analysis vividly demonstrates this aspect in terms of subchain
linear dimensions, with which we may associate the effective Flory
exponent which is clearly largest in the immediate vicinity of the
$cis$ side of the pore, and acquires a smaller value in the in the
immediate vicinity of the $trans$ side of the pore. Immediately
after the translocation process clearly the entire chain is
compressed and described by a lower value of the effective Flory
exponent. On the contrary the $cis$-segments are described by a
progressively higher value of the effective Flory exponent. In
short, the bias at the pore makes the chain expanded and
compressed at the $cis$ and $trans$ sides respectively, the chain
tension being a function of the location of the segment (bead)
with respect to the pore. This aspect was emphasized by Sakaue in
a slightly different context. From the chain segment analysis we
find that the translocation process can be viewed as a propagation
of a ``defect'' in the value of the effective Flory exponent space
in the opposite direction of translocation. Therefore, the
analytic treatments in terms of one ``slow'' variable ($s$
coordinate) is not adequate. Since the relaxation is different on
the $cis$ and $trans$ side, one needs to incorporate additional
slow variables to account for this $trans-cis$
asymmetry\cite{Sung}. We also present some elements of a phenomenological 
theoretical description of nonequilibrium behavior, in terms of an effective 
time-dependent correlation length that describes on which scales 
excluded volume interactions during translocation.\par
With the propagating defect picture in mind let us now look at the
possible scenario in the limit of very large chain. In this limit 
Fig.~\ref{subchain_nu} will look like as shown in Fig.~\ref{defect}, reaching 
the equilibrium value at each side farthest from the pore.
How does the magnitude of the force dictate the correlation length
of the decay of this defect ? Furthermore, we know that in the
limit of force $\rightarrow 0$ the segment analysis will remain
valid and the defect will go to zero as there will be no
distinction between the $trans$ and $cis$ side. Can one recover
Kantor-Kardar result\cite{Kantor04} from a finite size chain simulation with
proper extrapolation ? 
While further work is needed to settle these issues we have initiated 
some work along this line in section-III-H. 
We hope that the qualitative description of
polymer translocation in terms of a ``defect'' in the value of the
effective Flory exponent will motivate further theoretical work to
construct a more complete theory of forced translocation.
\section{ACKNOWLEDGEMENT}
A. B. appreciates the local hospitality at the Institut f\"{u}r
Physik, Johannes-Gutenberg Universit\"{a}t, Mainz, where most of
this work was done, and gratefully acknowledges the travel support
from the Deutsche Forschungsgemeinschaft, SFB 625/A3 and from the
Schwerpunkt f\"ur Rechnergest\"utzte Forschung in den
Naturwissenschaften (SRFN). We also thank both the referees for their 
constructive comments.
\par \noindent\medskip \vskip 1.0truecm \centerline{REFERENCES}

\newpage
\centerline{FIGURE CAPTIONS}
\noindent{\bf Fig. 1:}
Scaling plot (log-log scale) of the mean
translocation time $\langle \tau \rangle$ for the case where no
bias force $F$ is applied, as a function of chain length $N$. The
inset shows the corresponding plots for the gyration radii where the 
diamonds, circles, squares, represent the gyration radii, and the  
longitudinal and the transverse components $R_{gl}$ and 
$R_{gt}$ respectively. \par
\medskip\noindent{\bf Fig. 2:}~
(a) Mean first passage time (MFPT) for
forced translocating chains. 
Different colors correspond to  
N=16(black), N=32(red),
N=64(green), N=128(blue) and N=256(orange). (b) MFPT normalized by
the $\langle \tau(N) \rangle$ as a function of normalized chain
segment $m/N$ for N=16(top curve) to N=256(bottom curve). 
The dashed straight line (magenta) corresponds to
the line of unit slope for comparison\par

\medskip\noindent{\bf Fig. 3:}~
Location of the $Z_{CM}(N/2)$ for
several different chain lengths plotted vs. the index labelling
the individual simulation runs. The dots represent data points
from each run. The dashed red line correspond to the $\langle
Z_{CM}(N/2) \rangle $ averaged over all these points. The position
of $Z=0$ is denoted by the solid black line for easy comparison.\par

\medskip\noindent{\bf Fig. 4:}~
Plot of $Z_{CM}(N/2)$ as a function of chain length for 
several different bias values. The circles, squares, diamonds, and triangles 
correspond to the bias values F=12.0, F=6.0, F=4.0, and F=2.0 respectively.\par
\medskip\noindent{\bf Fig. 5:}~
(a) Plot of $\langle Z_{CM}^{cis}(n)\rangle$ (dashed black, red and green) and 
$\langle Z_{CM}^{trans}(n) \rangle$ (solid blue, orange, and magenta) as 
a function of the segments on the $cis$ and $trans$ side for chain lengths 
N = 65, 129 and 257 respectively for unbiased translocation. 
(b) the corresponding plot on the log-log scale where we notice that for a large fraction
of the $cis$ and $trans$ segments $\langle Z_{CM}(n) \rangle \sim m^{0.57}$.\par
\medskip\noindent{\bf Fig. 6:}~
(a) Plot of $\langle Z_{CM}^{cis}(n)\rangle$ (black, red and green) and
$\langle Z_{CM}^{trans}(n) \rangle$ (blue, orange, and magenta) as
a function of the segments on the $cis$ and $trans$ side for chain lengths 
N=64, 128, and 256
respectively for forced translocation. Note that the
data for $\langle Z_{cm}^{trans} (n)\rangle $ for different $N$
coincide almost perfectly on a common curve, so one can hardly
distinguish the individual values for $m < 64$ on this plot;
(b) the corresponding plots on log-log scale. Straight lines indicate
the power laws quoted in the figure. \par

\medskip\noindent{\bf Fig. 7:}~
Plot of gyration radii $\langle R_g(t) \rangle $ as a function of time for the unbiased
(top) translocation and (bottom) for the forced translocation where a significant variation of the
size of the translocating chain is immediately noticeable. In each figure that double-dotted 
(black and blue) lines represent the longitudinal components, wide dashed (red and orange) lines
represent the transverse components and the solid lines (green and magenta) correspond to the gyration 
radii for chain lengths 65(64) and 129(128) for the unbiased(biased) translocation respectively. The thin straight lines in each case (with the same symbols) correspond to the corresponding time averaged values. Note that 
 $\langle R_g(t) \rangle > \langle R_{gl}(t) \rangle >  \langle R_{gt}(t) \rangle$
for unbiased translocation.

\par
\medskip\noindent{\bf Fig. 8:}~
Plot of (log-scale) (a) $\langle
R_{gt}^{cis}(n) \rangle$, $\langle R_{gt}^{trans}(n) \rangle$, (b)
$\langle R_{gl}^{cis}(n)$, $R_{gl}^{trans}(n) \rangle$, and (c)
$\langle R_{g}^{cis}(n) \rangle$, $ \langle R_{g}^{trans}(n)
\rangle $ as a function of the segments $m$ on the $cis$ and
$trans$ sides respectively, for the case of unbiased
translocation. The straight line (purple) in each figure refers to the slope. The inset of (b) compares the longitudinal component 
(upper curve) $\langle R_{gl}(n)\rangle $ and the transverse component (lower curve)
$\langle R_{gt}(n)\rangle $ for chain length N=129 for
comparison.\par
\medskip\noindent{\bf Fig. 9:}~
{Plot of (log-scale) (a) $\langle R_{gt}^{cis}(n) \rangle$,
$\langle R_{gt}^{trans}(n) \rangle$,
(b) $\langle R_{gl}^{cis}(n)$, $R_{gl}^{trans}(n) \rangle$, and
(c) $\langle R_{g}^{cis}(n) \rangle$, $ \langle R_{g}^{trans}(n) \rangle $
as a function of the segments $m$ on the
$cis$ and $trans$ sides respectively. In each figure the solid and dotted lines correspond to 
 the $cis$ and $trans$ components respectively.
Note that the curves for the $trans$ part (lower set of dashed curves
in each part of the figure) superimpose to such a large extent
that they are hardly distinguishable from the data for $N=256$ in
these figures.\par

\medskip\noindent{\bf Fig. 10:}~
Plot of the radius of gyration $\langle R_g \rangle$ and end
to end distance $\langle R_N \rangle$ (logarithmic scale) of the chain which has just translocated as a function of
the chain length N (logarithmic scale).  The solid(open) circles (black) and 
squares (red) refer to the forced(unbiased) translocation. \par

\medskip\noindent{\bf Fig. 11:}~
(a) Plot of longitudinal (solid lines) and transverse (dotted lines) components of gyration radii 
for the polymer which has just
translocated as a function of the monomer index $m$ for unbiased translocation. 
The solid black, red, and green lines correspond to the longitudinal components 
and dotted blue, orange, and magenta lines
correspond to the transverse components for chain lengths N=65 and N=129 respectively. 
(b) the plot on a log-log scale. \par

\medskip\noindent{\bf Fig. 12:}~
Plot of gyration radii for the chain segments of length $n=32$ for chain
length $N=256$ at different stages of the translocation process when integral number of
chain segments $n_i$ have translocated. The vertical red lines differentiate
the $trans$ and the $cis$ segments. The filled squares and diamonds represent the longitudinal and
the transverse components of the gyration radii respectively and the filled circles represent
the the gyration radii. \par

\medskip\noindent{\bf Fig. 13:}~
Plot of effective Flory exponents as a function of the subchain index at different 
stage of the translocation process. 
The filled squares and diamonds represent the effective exponents for the 
longitudinal and the transverse components of the gyration radii respectively 
and the filled circles represent the effective exponent for the gyration radii. \par

\medskip\noindent{\bf Fig. 14:}~
(a) Plot of longitudinal (solid lines) and transverse (dotted lines) gyration radii 
for the polymer which has just
translocated as a function of the monomer index $m$. The solid black, red, and green lines 
and dotted blue, orange, and magenta lines 
correspond to the chain lengths N=64, 128, and 256 for the longitudinal and transverse components
respectively. (b) the same plot on a log-log scale.\par

\medskip\noindent{\bf Fig. 15:}~
The distribution of longitudinal (circles) and 
transverse (squares) 
components of the end to end distances of a chain which has just
translocated for chain lengths N=128 (top) and N=256 (bottom). 
For comparison the corresponding equilibrium configurations for 
the longitudinal (diamonds) and transverse (triangles) 
components at the $cis$ side at the beginning of each translocation 
run are also shown.\par

\medskip\noindent{\bf Fig. 16:}~
Contemplated defect schematic in the vicinity of the pore.
The range of segment number
$m$ over which a change of the effective exponent is observed
increases with increasing size of the subchain that is
considered. \par
\newpage

\begin{figure}[ht!]
\begin{center}
\includegraphics[width=12truecm]{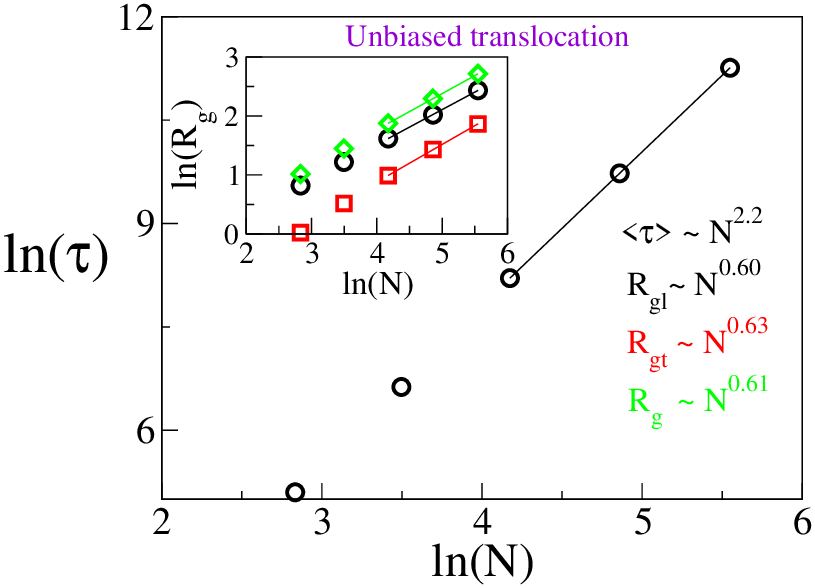}
\vskip 3.0truecm
\caption{}
\label{tauf0}       
\end{center}
\end{figure}
\begin{figure}[ht!]
\begin{center}
\includegraphics[width=12truecm]{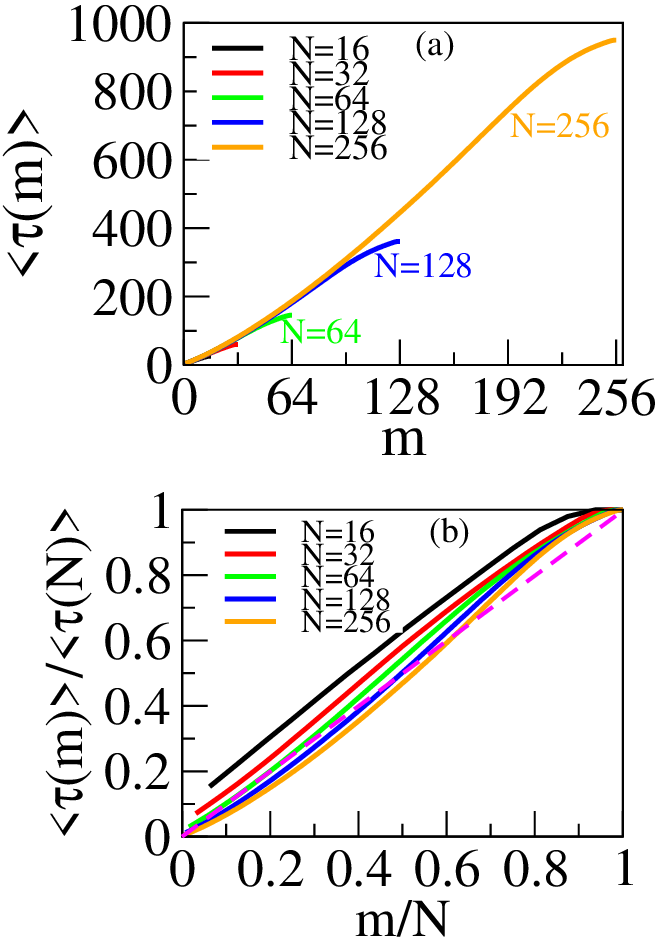}
\vskip 2.0truecm
\caption{}
\label{mfpt}       
\end{center}
\end{figure}
\begin{figure}[ht!]
\begin{center}
\includegraphics[width=10truecm]{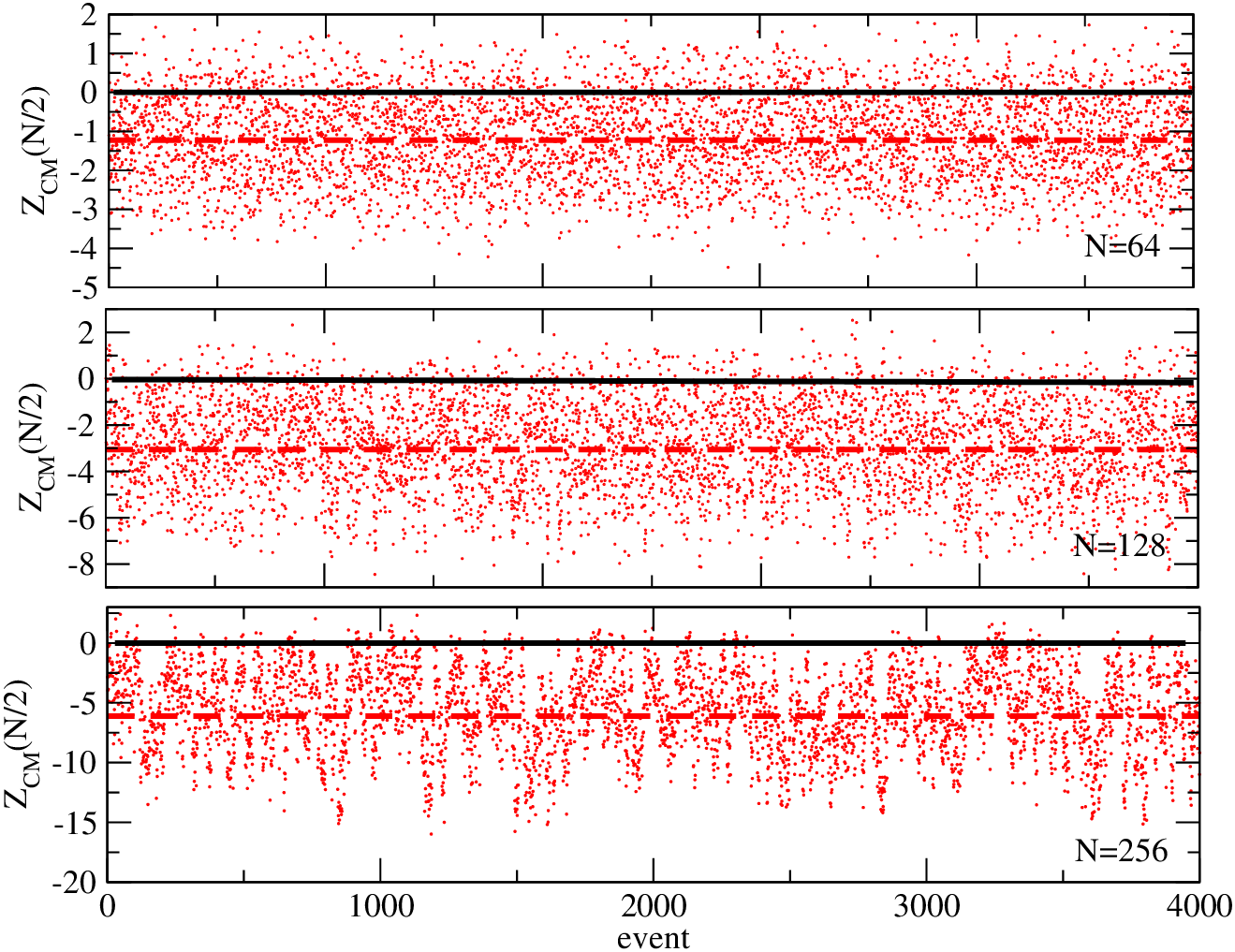}
\vskip 3.0truecm
\caption{}
\label{cmznh_t}       
\end{center}
\end{figure}
\begin{figure}[ht!]
\begin{center}
\vskip 0.50truecm
\includegraphics[width=10truecm]{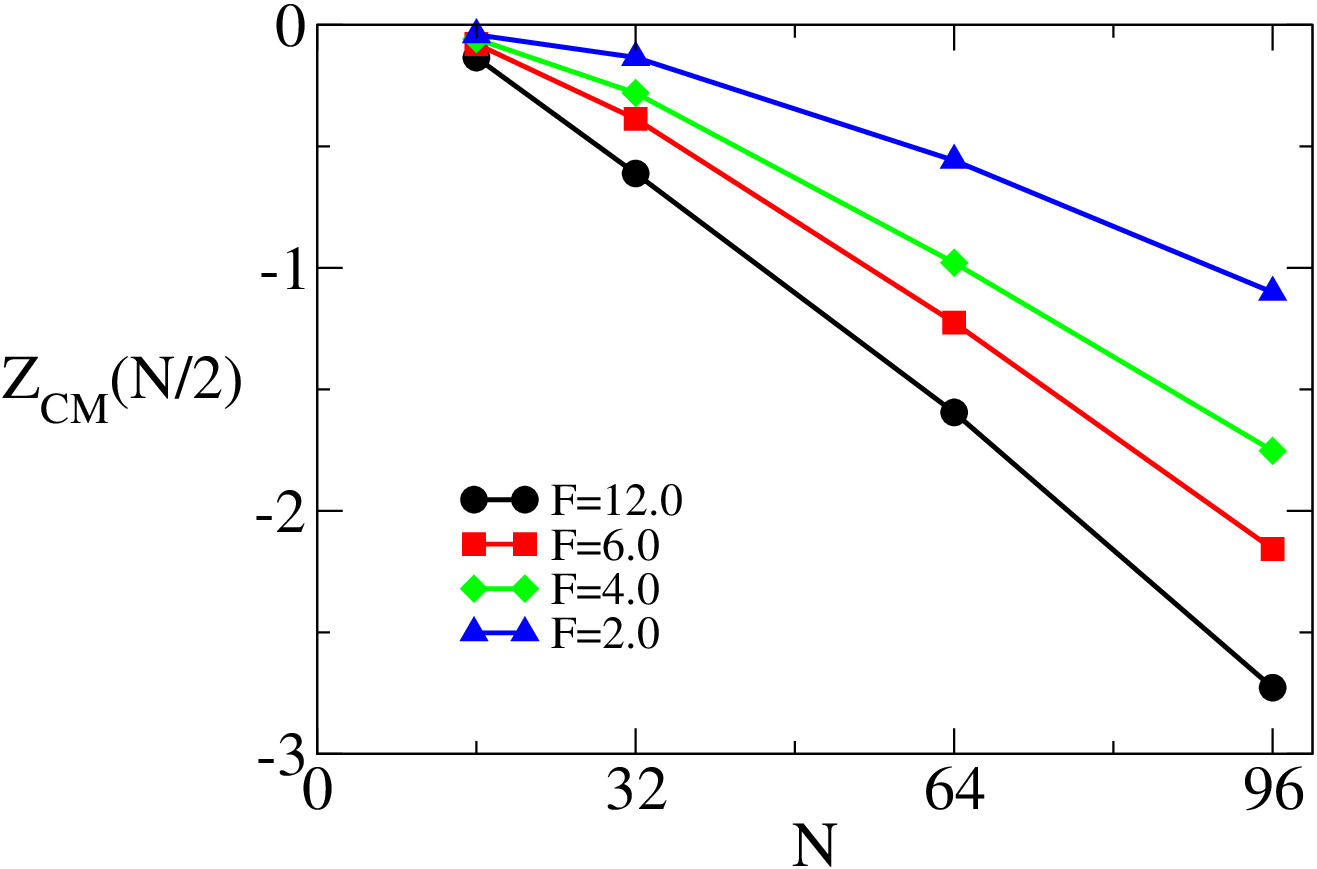}
\vskip 2.0truecm
\caption{}
\label{cmznh_N}       
\end{center}
\end{figure}
\begin{figure}[ht!]
\begin{center}
\includegraphics[width=12truecm]{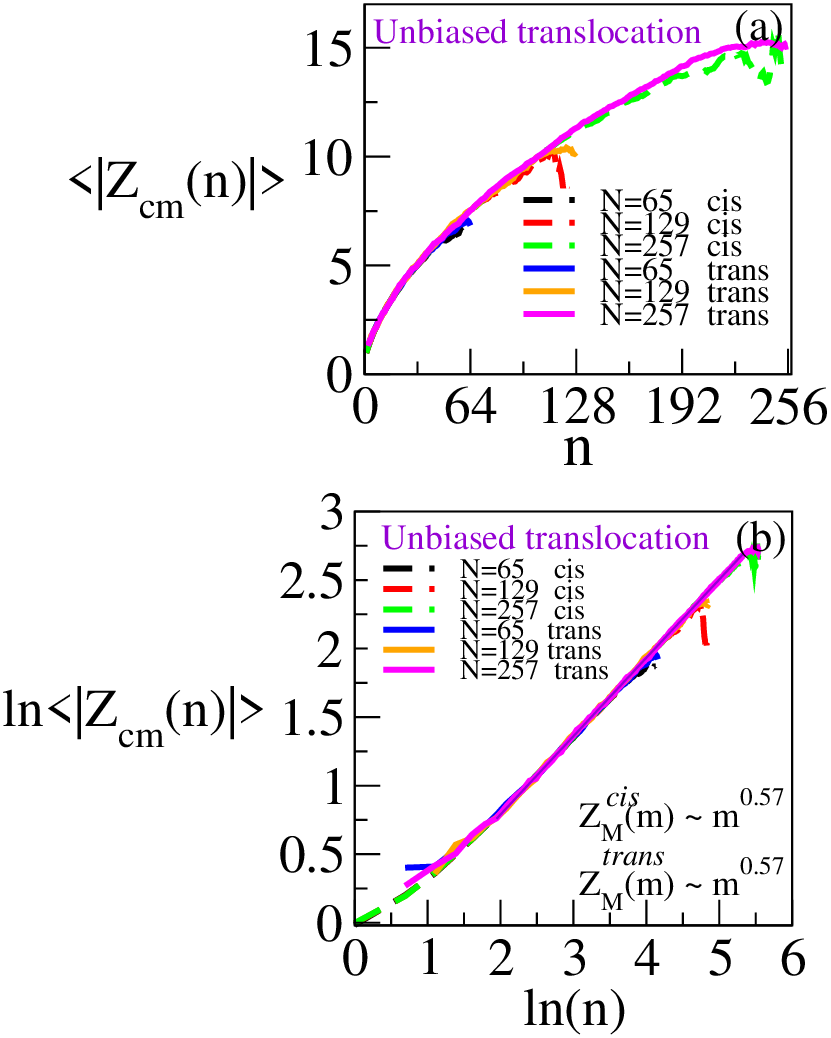}
\vskip 2.0truecm
\caption{}
\label{cmzs_unbiased}       
\end{center}
\end{figure}
\begin{figure}[ht!]
\begin{center}
\includegraphics[width=12truecm]{cmzs_forced.eps}
\vskip 2.0truecm
\caption{}
\label{cmzs_forced}       
\end{center}
\end{figure}
\begin{figure}[ht!]
\begin{center}
\includegraphics[width=12truecm,angle=270]{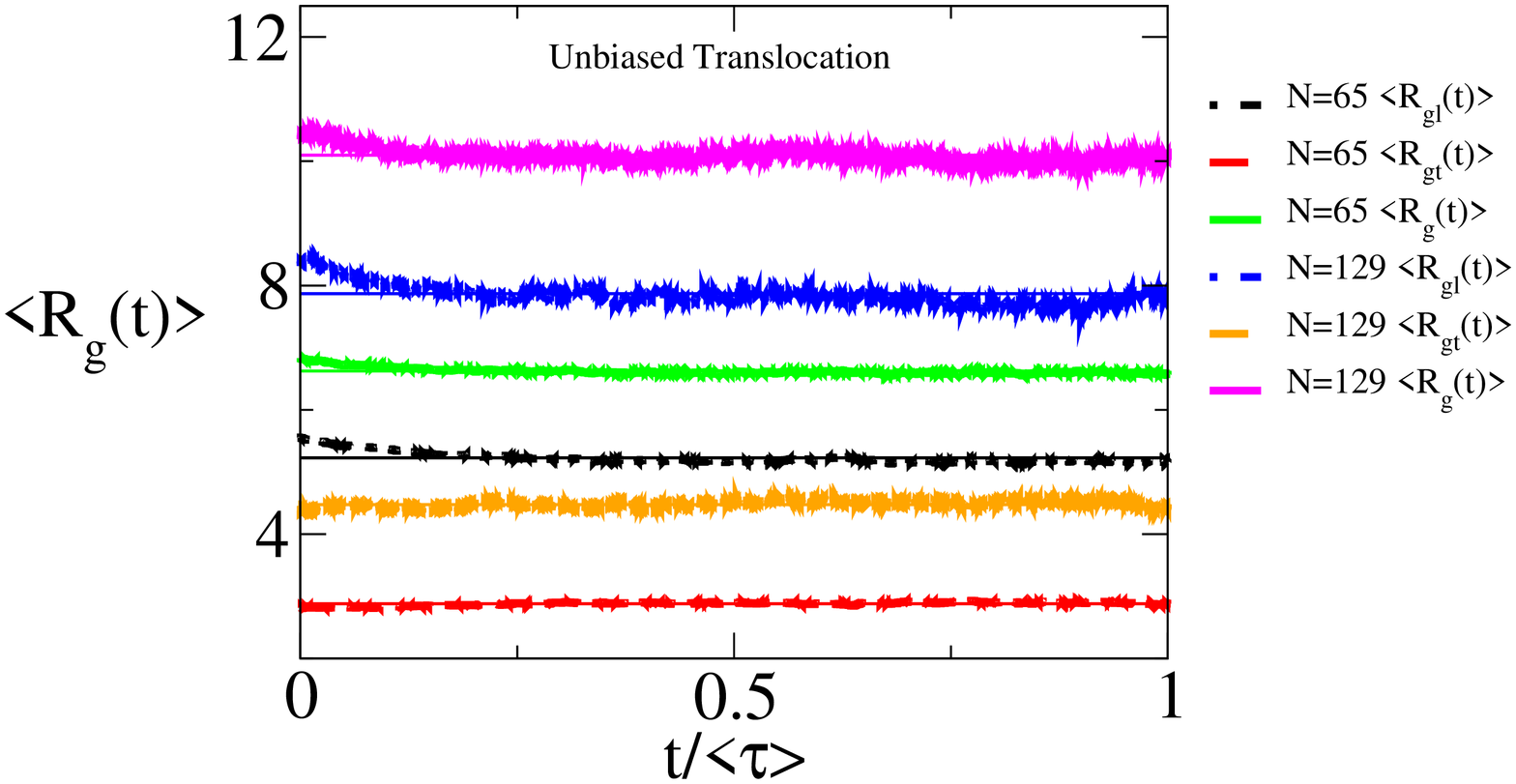}
\includegraphics[width=12truecm]{rgt_forced.eps}
\vskip 2.0truecm
\caption{}
\label{rgt}       
\end{center}
\end{figure}
\begin{figure}[ht!]
\begin{center}
\includegraphics[width=14truecm]{rgs_unbiased.eps}
\vskip 3.0truecm
\caption{}
\label{rgs_unbiased}       
\end{center}
\end{figure}
\begin{figure}[ht!]
\begin{center}
\includegraphics[width=10.0truecm]{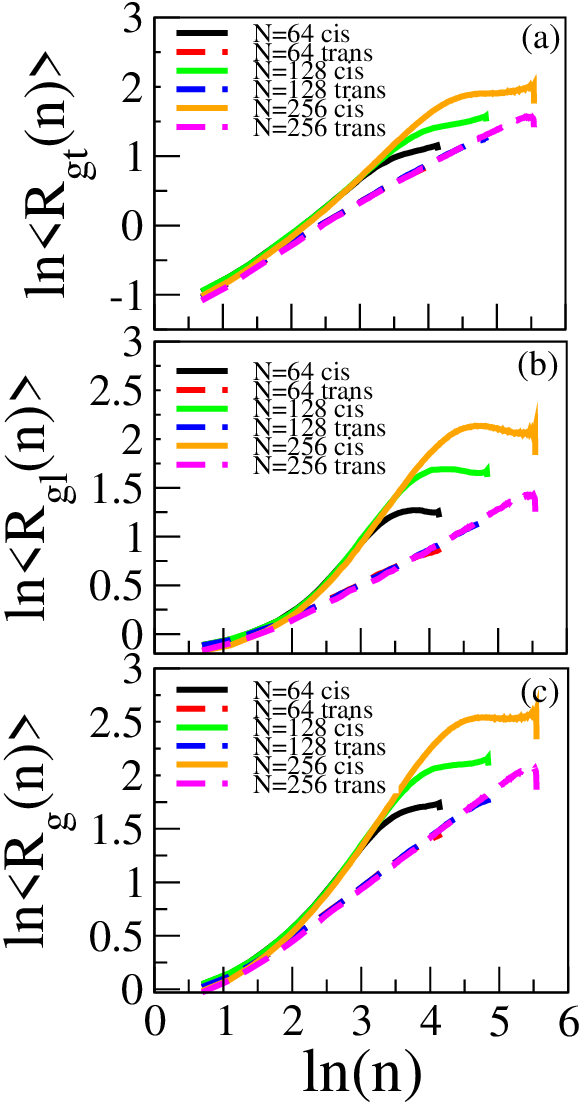}
\vskip 3.0truecm
\caption{}
\label{rgs_forced}
\end{center}
\end{figure}
\begin{figure}[ht!]
\begin{center}
\includegraphics[width=14truecm]{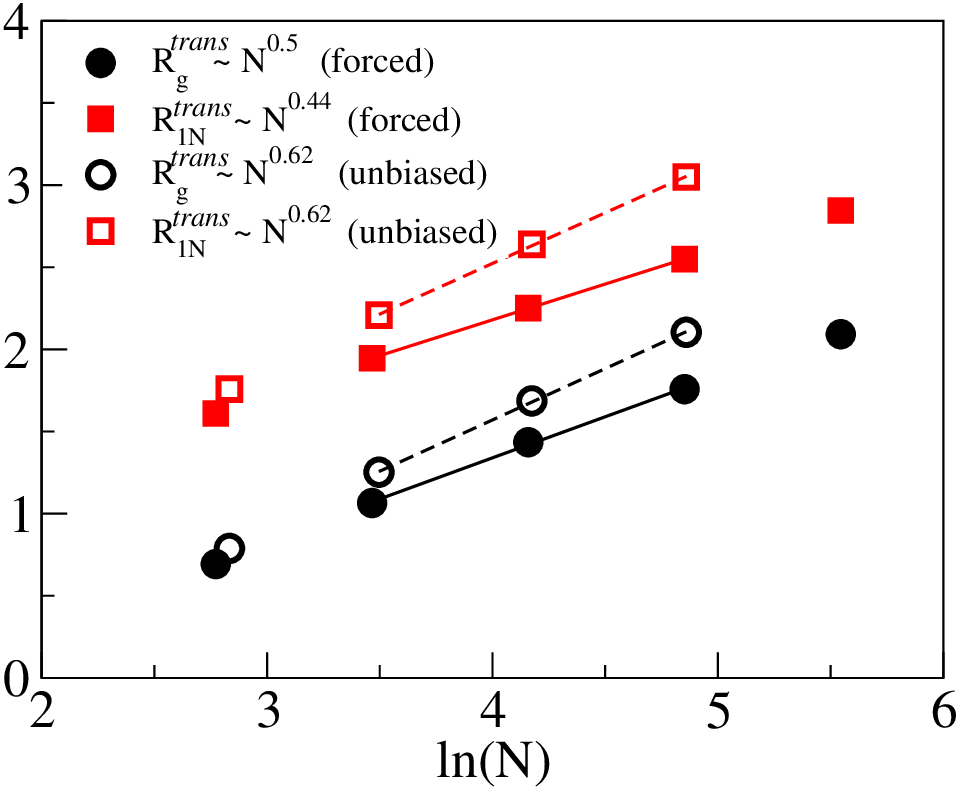}
\vskip 3.0truecm
\caption{}
\label{rg_r1n_trans}       
\end{center}
\end{figure}
\begin{figure}[ht]!
\begin{center}
\includegraphics[width=12truecm]{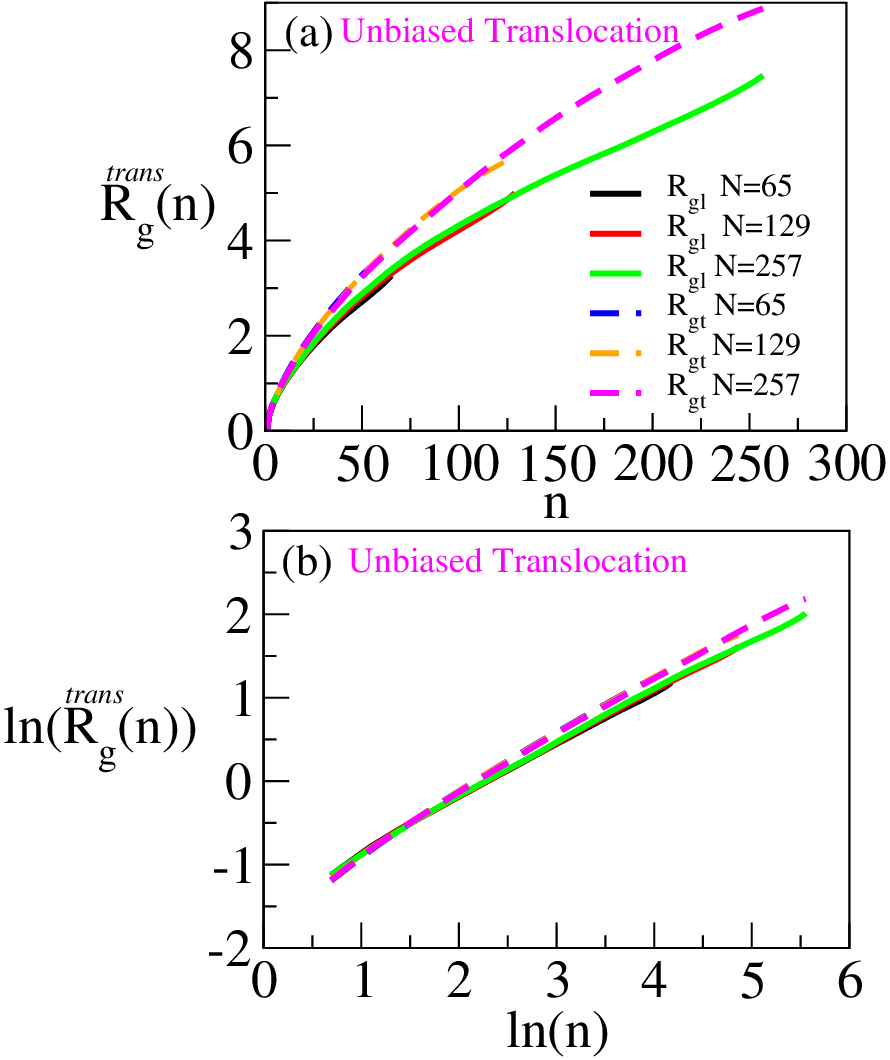}
\vskip 3.0truecm
\caption{}
\label{rgtrans_unbiased}       
\end{center}
\end{figure}
\newpage
\begin{figure}[ht!]
\begin{center}
\includegraphics[width=15truecm,angle=0]{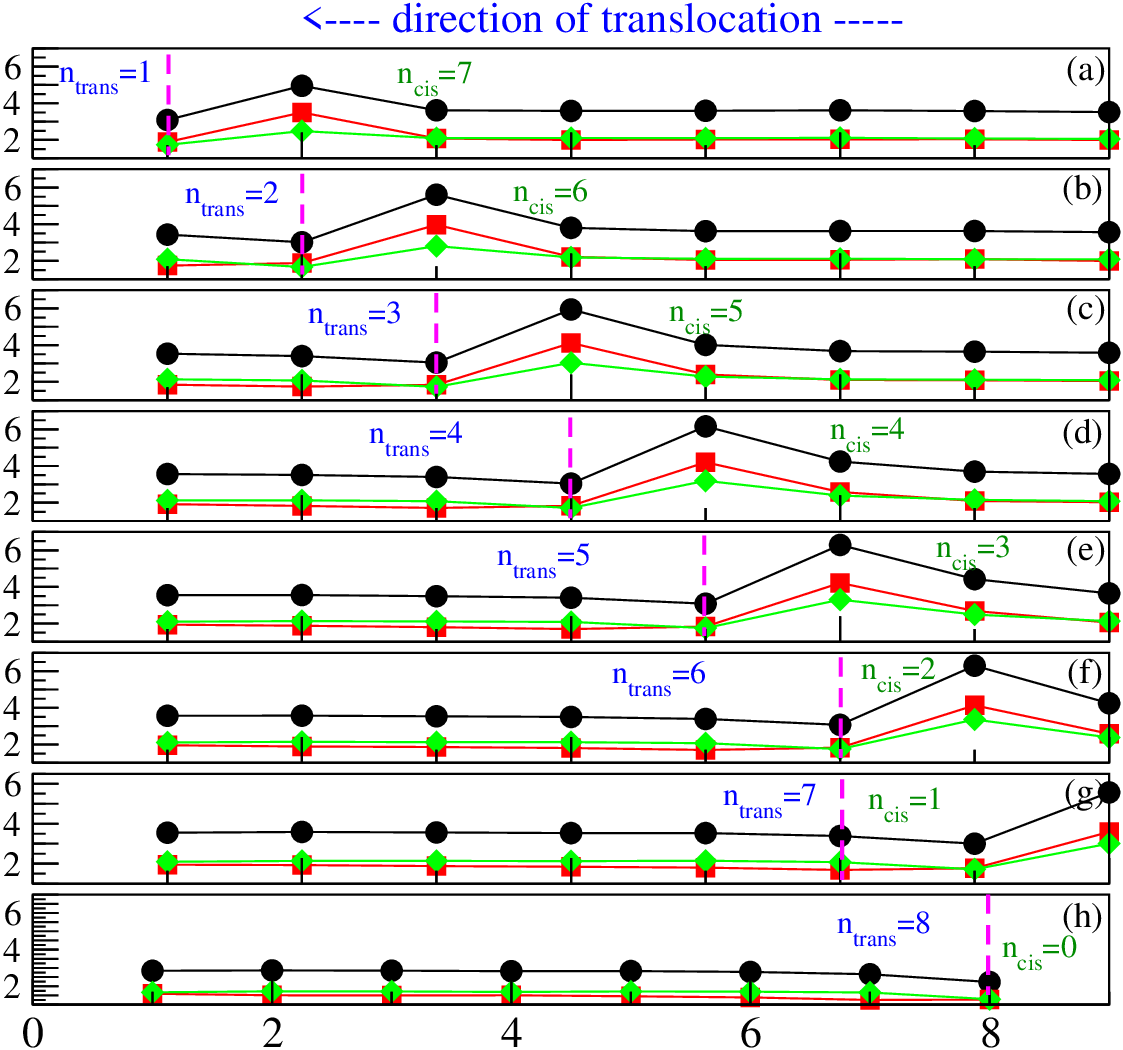}
\vskip 3.0truecm
\caption{}
\label{subchain32}       
\end{center}
\end{figure}
\begin{figure}[ht!]
\begin{center}
\includegraphics[width=15truecm,angle=0]{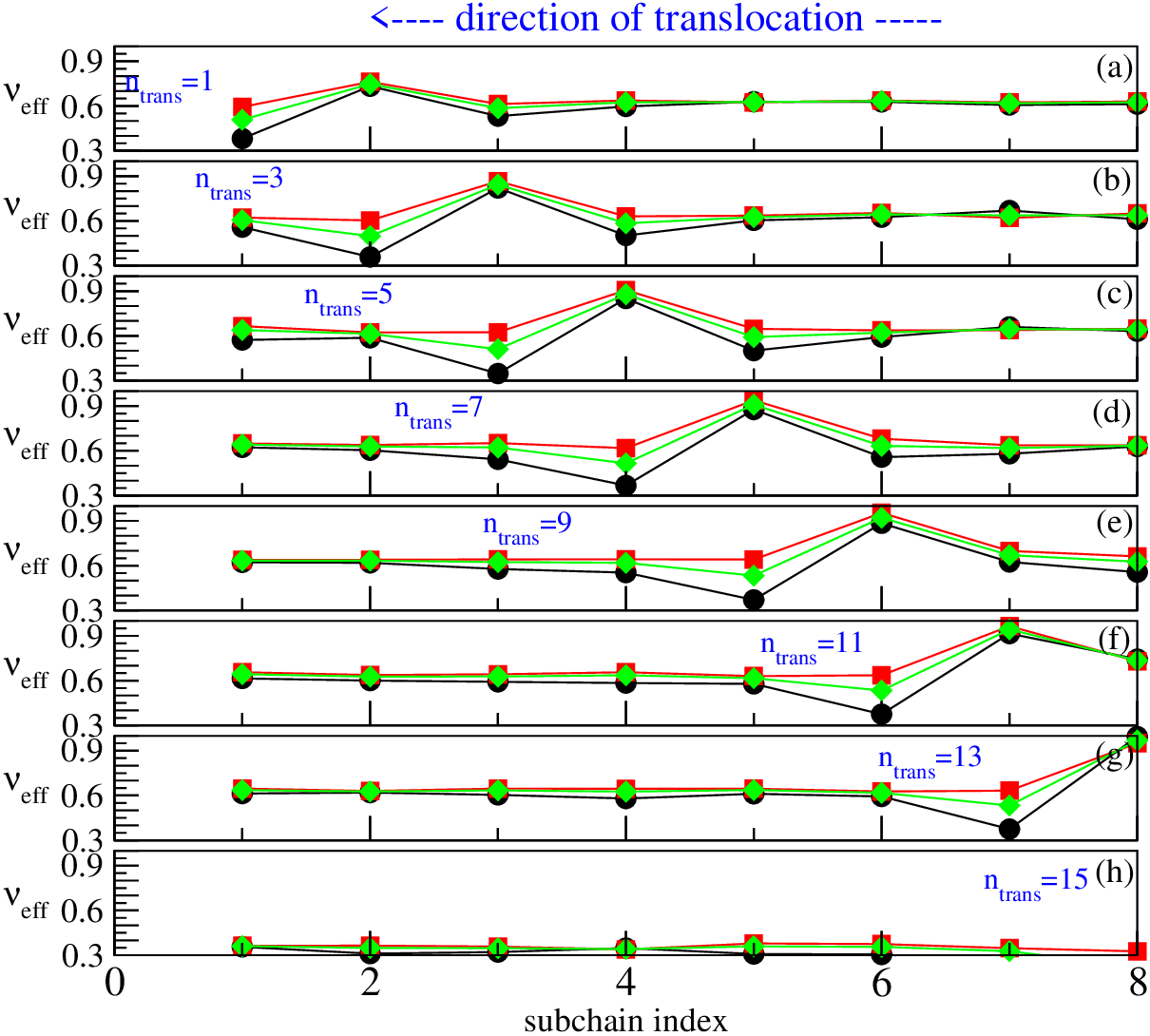}
\vskip 3.0truecm
\caption{}
\label{subchain_nu}       
\end{center}
\end{figure}
\newpage
\begin{figure}[ht!]
\begin{center}
\includegraphics[width=14truecm]{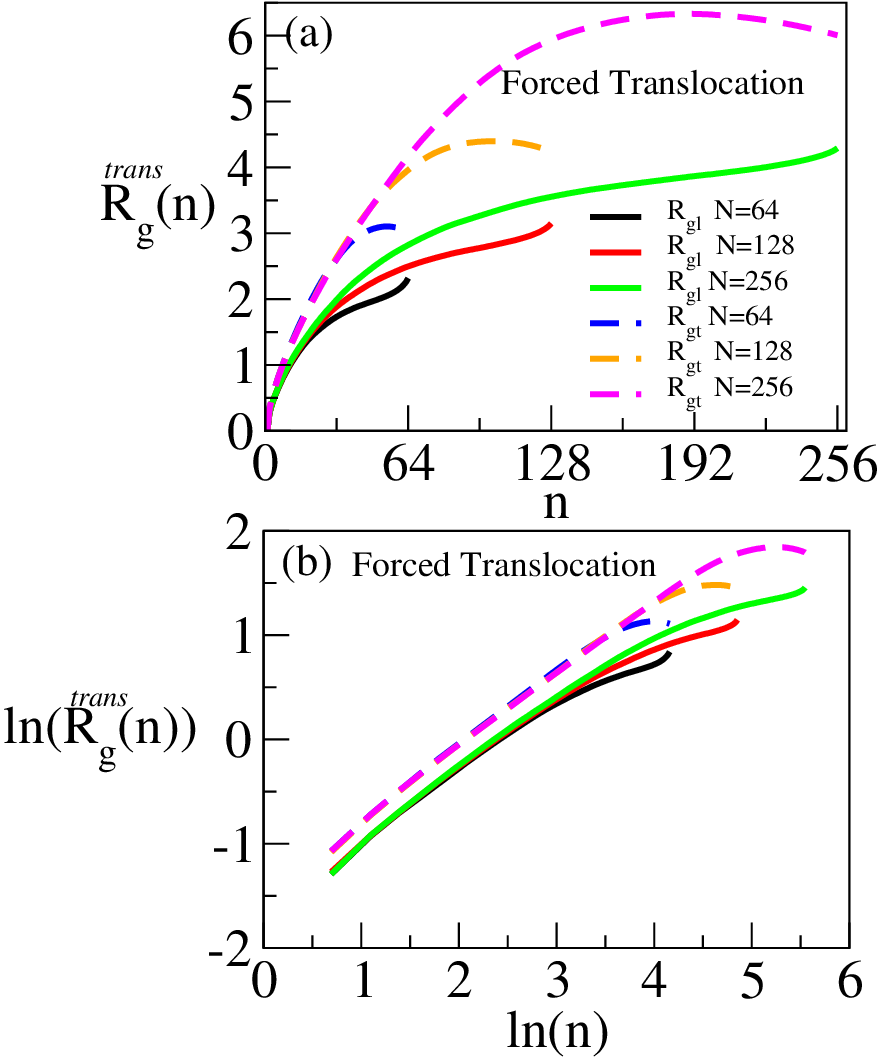}
\vskip 3.0truecm
\caption{}
\label{rgtrans_forced}       
\end{center}
\end{figure}
\begin{figure}[ht!]
\begin{center}
\includegraphics[width=14truecm]{r1n_trans_dist.eps}
\vskip 2.0truecm
\caption{}
\label{r1n_hist}       
\end{center}
\end{figure}
\begin{figure}[ht!]
\begin{center}
\includegraphics[width=14truecm]{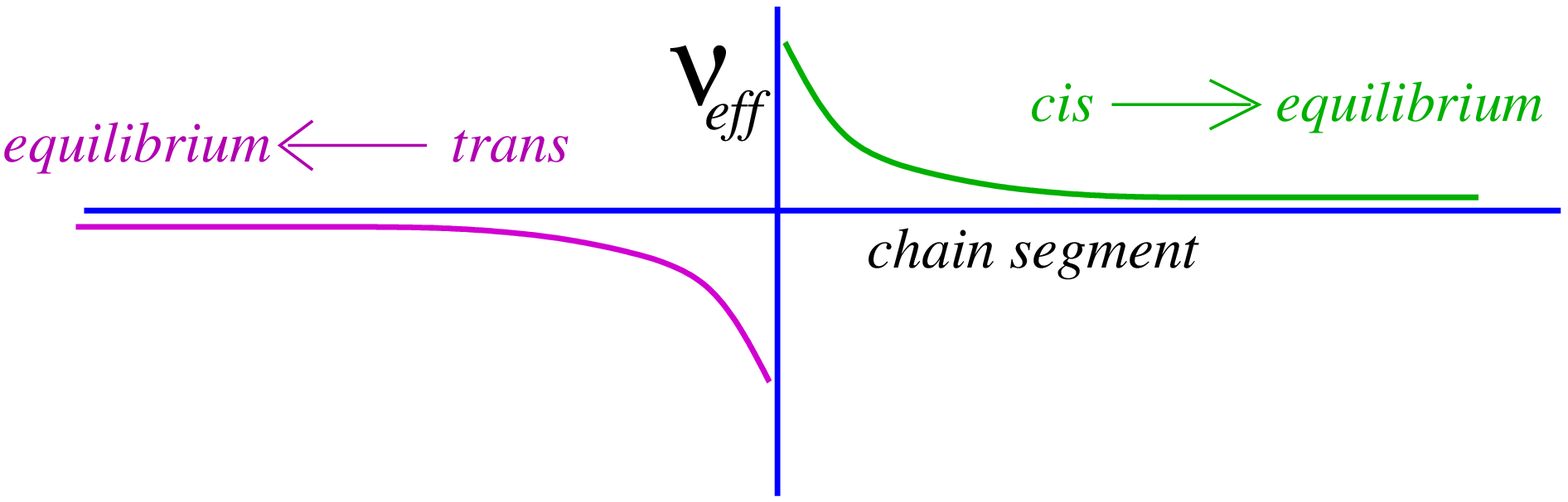}
\vskip 3.0truecm
\caption{}
\label{defect}       
\end{center}
\end{figure}

\begin{thebibliography}{99}
\bibitem{comment1} For the case of unbiased translocation the $cis$/$trans$ components
are defined if the chain translocates from left($cis$) to right($trans$) or
vice versa unlike the case of forced translocation where the applied bias uniquely
defines the $cis$ and the $trans$ components.
\end{thebibliography}
\end{document}